\documentstyle[prd,floats,psfig,aps]{revtex}

\begin{document}

\title{Evolution of Kerr-Schild type initial data for binary
black holes using the horizon penetrating Teukolsky equation}

\author{Gaurav Khanna}
\address{Natural Science Division,\\
Southampton College of Long Island University,\\
Southampton NY 11968.}

\maketitle

\begin{abstract}
We use the Kerr-Schild type Teukolsky equation (horizon penetrating) to evolve
binary black hole initial data as proposed by Bishop {\em et al.} in the close limit. 
Our results are in agreement with those recently obtained by Sarbach {\em et al.} 
from the Zerilli equation evolution of the same initial data. 
\end{abstract}

\section{Introduction}

There is a lot of current interest in studying the collision of two black holes since these events 
could be primary sources of gravitational waves for interferometric gravitational wave detectors 
currently under construction. Numerical simulations are considered as the key to model such collisions. 
Unfortunately, current computer limitations cause these simulations to be short lived, so  
to study the actual merger of two holes, one needs to have initial data that represents two black holes 
close to each other. 

Early attempts to provide binary black hole initial data  solved the initial value 
problem of General Relativity using conformally flat slices \cite{C}.  This restriction was too 
strong; it was unable to incorporate spinning holes. Lately, 
some effort has been geared towards obtaining initial data based on the
Kerr--Schild form of black hole solutions. The main advantage of this form is that the spatial
slices of these solutions are horizon penetrating, which make them ideal for the application 
of the ``excision'' method \cite{maya} for evolving black holes. This form of the solution also 
naturally incorporates boosted and spinning black holes.

There is also another way \cite{KL} to obtain initial data for a black hole binary, without solving 
the initial value equations directly. One could, in principle, consider a scenario with matter fields
coupled to gravity such that there are no black holes to begin with. One could distribute the
matter fields in such a way that their gravitational effects are like those of two stars that are 
separated far apart but each is critically close to gravitational collapse. It is quite possible 
that such a simulation would  not be plagued with the subtle problems that  ocurr when one attempts 
to evolve two black holes. One could set the parameters of the simulation in such a way that when the two stars
come close to each other, they collapse upon themselves and form two black holes. This final slice could be used
as initial data for a binary black hole code. 

However in this paper, we consider the Bishop {\em et al.} \cite{BIMW} family of initial data 
wherein two non spinning Kerr--Schild black holes are superposed.  This  proposal has the
 property that in the ``close limit'' in which the separation of the holes is small, the
metric is given by a distorted Kerr-Schild black hole.  The evolution will be carried out 
using a recently introduced horizon penetrating Teukolsky formalism \cite{CKLPR}.  We 
present Teukolsky waveforms from this evolution and the amount of energy radiated 
away by the system. The results we obtain are in agreement with those obtained by 
Sarbach {\em et al.} \cite{STP} from their evolution of the same initial data family using a Kerr--Schild  
type Zerilli equation.

\section{Kerr--Schild initial data}

For the sake of completion, in the first subsection of this section we reproduce
the calculations followed by Sarbach {\em et al.} \cite{STP} for an exact solution of the 
linearized initial value problem. 

The KS  space-time metric is defined by
\begin{displaymath}
g_{\mu\nu} = \eta_{\mu\nu} - 2V k_\mu k_\nu\, ,
\end{displaymath}
where $k_\mu$ is a null vector.
The  $\eta_{\mu\nu}$ is taken to be the Minkowski metric with coordinates 
$(t,\underline{x}) = (t,x^i)$ such that $\eta_{tt} = -1$, $\eta_{ti} = 0$ and $\eta_{ij} = \delta_{ij}$.
The null vector $k_\mu$ satisfies $k_t = -1$ and
$k^i k_i = 1$, where $k^i = \delta^{ij} k_j$. The three metric and extrinsic curvature
with respect to a slice $t = \mbox{const.}$ are
\begin{eqnarray}
\bar{g}_{ij} &=& \delta_{ij} - 2V k_i k_j,
\label{Eq-KSgK1}\\
K_{ij} &=& -\frac{1}{\alpha}\partial_t \left( V k_i k_j \right)
 + 2\alpha \left[ V k^s \nabla_s \left( V k_i k_j \right) -
 \nabla_{(i} \left( V k_{j)} \right) \right],
\label{Eq-KSgK2}
\end{eqnarray}
where $\alpha = (1 - 2V)^{-1/2}$ is the lapse and where
$\nabla$ refers to the flat metric $\delta_{ij}\,$.

Bishop {\em et al.}'s solution procedure \cite{BIMW}  consists in inserting
equations \ref{Eq-KSgK1} and \ref{Eq-KSgK2} into the constraint equations
and to solve the resulting equations for $V$, $\dot{V} = \partial_t V$
and $\dot{k}_i = \partial_t k_i$, where $k_i$ is assumed to be given.

\subsection{Two black hole data}

Here we review the discussion of equations \ref{Eq-KSgK1} and \ref{Eq-KSgK2} as followed by
Sarbach {\em et al.} where a particular ansatz is made for $k_i$ representing two nearby
non-rotating and non-spinning black holes. A single Schwarzschild black hole can be written as
\begin{displaymath}
k_i = \frac{\nabla_i \phi}{|\nabla\phi|}\, , \;\;\;
|\nabla\phi|^2 = \delta^{ij} \nabla_i\phi\cdot\nabla_j\phi,
\end{displaymath}
with $\phi = 1/r$.
For two black holes, Bishop {\em et al.} choose 
\begin{displaymath}
\phi(\underline{x}) = \frac{M_1}{ |\underline{x} -
\underline{x}_1 | } + \frac{M_2}{ |\underline{x} - \underline{x}_2 | }\, ,
\end{displaymath}
where $\underline{x}_{1}$ and $\underline{x}_{2}$ are the positions of the black holes 
that have mass $M_{1}$ and  $M_{2}$.
If the black holes are located at $\underline{x}_1= (0,0,{a_1})$ and  
$\underline{x}_2= (0,0,{a_2})$, with $a_1 > 0 > a_2$, $\phi$ may be 
expanded in a sum over multipoles:
\begin{equation}
\phi(r,\vartheta) = \sum\limits_{\ell=0}^{\infty}
\frac{ M_1 a_1^\ell + M_2 a_2^\ell}{r^{\ell+1}} P_\ell(\cos\vartheta),
\label{Eq-TwoBHExp}
\end{equation}
where $P_\ell$ denote standard Legendre polynomials and where
$(r,\vartheta,\varphi)$ are polar coordinates for $\underline{x}$.
Note that the expansion (\ref{Eq-TwoBHExp}) is only valid for $r > \max\{a_1,-a_2\}$.
We define the separation parameter as
\begin{displaymath}
\varepsilon = \frac{a_1 - a_2}{M}\, ,
\end{displaymath}
where $M = M_1 + M_2$ is the total mass. Now, using 
 the center of mass condition $M_1 a_1 + M_2 a_2 = 0$,
the close limit of (\ref{Eq-TwoBHExp}) becomes
\begin{displaymath}
\phi = \frac{M}{r} + \varepsilon^2\frac{ M M_1 M_2}{r^3}
P_2(\cos\vartheta) + {\cal O}(\varepsilon^3/r^4).
\end{displaymath}
As a result, to first order in $\varepsilon^2$, $k_i$ is given by
\begin{equation}
k_r = -1, \;\;\;
k_A = \varepsilon^2\frac{M_1 M_2}{r} \hat{\nabla}_A P_2\, ,
\label{Eq-Exp1}
\end{equation}
where here and in the following, $A = \vartheta,\varphi$.
The remaining amplitudes are expanded according to
\begin{equation}
V = -\frac{M}{r} + \varepsilon^2 v(r) P_2\, , \;\;\;
\dot{V} = \varepsilon^2\dot{v}(r) P_2\, , \;\;\;
\dot{k}_A = \varepsilon^2\dot{k}(r) \hat{\nabla}_A P_2\, .
\label{Eq-Exp2}
\end{equation}
Plugging this into the constraint equations, and keeping only linear
terms (of the order $\varepsilon^2$), one obtains the equations
\begin{eqnarray}
0 &=& -\dot{v} + \frac{3M}{r^2}\left( 1 + \frac{2M}{r} \right)\dot{k}
      - \frac{3}{r}v - \frac{6M M_1 M_2}{r^5}(r - M),\label{Eq-C1}\\
0 &=& -v' - \frac{4}{r} v + \frac{6M^2}{r^3} \dot{k}
      - \frac{6M M_1 M_2}{r^5}(r - M), \label{Eq-C2}\\
0 &=& - M \dot{k}' + v + \frac{2M}{r} \dot{k} + \frac{6M M_1 M_2}{r^3}\, .
\label{Eq-C3}
\end{eqnarray}
Here, a prime denotes differentiation with respect to $r$.
The system (\ref{Eq-C2},\ref{Eq-C3}) can be re-expressed as
a single second order equation. Introducing the dimensionless
quantities $x = r/M$ and $\mu = M_1 M_2/M^2$, this equation reads
\begin{equation}
0 = -v_{xx} - \frac{5}{x} v_x + \frac{6}{x^3} v + \frac{6\mu}{x^6}(3x + 2).
\label{Eq-KeyEq}
\end{equation}
Once we have solved this equation, the remaining amplitudes
$\dot{k}$ and $\dot{v}$ are obtained from (\ref{Eq-C1}) and
(\ref{Eq-C2}), respectively.\\
A particular solution of (\ref{Eq-KeyEq}) is given by
\begin{displaymath}
v(x) = -\frac{2\mu}{3}\left( 1 - \frac{2}{x} +
\frac{3}{x^2} + \frac{3}{x^3} \right).
\end{displaymath}
In order to find the solutions of the homogeneous equation, one
performs the transformations $x = 24/z^2$, $v(x) = z^4 u(z)$,
which yields the Bessel differential equation
\begin{displaymath}
0 = z^2 u_{zz} + z u_z - (16 + z^2) u.
\end{displaymath}
The solutions are a linear combination of the Bessel functions
$I_4(z)$ and $K_4(z)$. 
Thus, the general solution to (\ref{Eq-KeyEq}) is $v(x) = \mu \hat{v}(x)$, with
\begin{equation}
\hat{v}(x) =  -\frac{2}{3}\left( 1 - \frac{2}{x} +
\frac{3}{x^2} + \frac{3}{x^3} + \frac{A_1}{x^2} K_4( \sqrt{24/x} ) \right)
     + \frac{A_2}{x^2} I_4( \sqrt{24/x} ) .
\label{Eq-vSol}
\end{equation}
Choices for $A_1$ and $A_2$ are discussed in the following section. 

\subsection{Initial data for the Teukolsky function}

The initial data for the metric and extrinsic curvature obtained above will be evolved using 
the KS type, penetrating Teukolsky equation. This equation was derived by Campanelli {\em et al.} 
\cite{CKLPR} in order to study gravitational perturbations close to the horizon. The equation, 
along with the associated geometrical quantity, the Newman-Penrose complex null-tetrad is regular at 
the horizon, and therefore has the capability of evolving through it. Such an equation is ideal to
study the effects of  ``excision'' as used by full numerical, binary black hole collision codes, 
like MAYA \cite{maya}. It should be noted, that the penetrating Teukolsky equation is based on a
rescaling of the transformed Kinnersley tertrad (to KS coordinates). This rescaling enables the 
tetrad to be regular at the horizon. Therefore, the penetrating Teukolsky equation is not just the 
original  Teukolsky equation in KS coordinates.\footnote{We recently learnt that Carsten 
Kollein at Albert Einstein  Institute performed a perturbative evolution of this same family of 
initial data in his M.Sc. thesis (unpublished), using the Teukolsky formalism. However, 
the Teukolsky equation used by Carsten was the Boyer-Lindquist Teukolsky equation transformed to 
KS coordinates, and not the penetrating Teukolsky equation. We thank Jorge Pullin and 
Manuela Campanelli for providing us with a copy of Carsten's work.}. 

To obtain initial data for the Teukolsky function, we use the formula developed by Campanelli 
{\em et al.} \cite{CaLoBaKhPu} in terms of the perturbative three metric and extrinsic curvature as 
shown below. $\psi_4$ is a Newman-Penrose scalar, that captures information about outgoing gravitational 
radiation, and it is related to the Teukolsky function $\psi$ as, $\psi\, =\,{r^4}{\psi_4}$ (for a 
non spinning black hole). 

\begin{eqnarray}
\psi _4 &=&-\left[ {R}_{ijkl}+2K_{i[k}K_{l]j}\right] _{(1)}n^i%
{m}^jn^k{m}^l+8\left[ K_{j[k,l]}+{%
\Gamma }_{j[k}^pK_{l]p}\right] _{(1)}n^{[0}{m}^{j]}n^k{m}^l
\label{psi} \\
&&\ -4\left[ {R}_{jl}-K_{jp}K_l^p+KK_{jl}
\right] _{(1)}n^{[0}{m}^{j]}n^{[0}{m}^{l]}, 
\nonumber\\
\partial _t\psi _4 &=&N_{(0)}^\phi \partial _\phi \left( \psi _4\right) -n^i%
{m}^jn^k{m}^l\left[ {\partial }_0R_{ijkl}\right]
_{(1)}  \label{psipunto} \\
&&+8n^{[0}{m}^{j]}n^k{m}^l\left[ {\partial }
_0K_{j[k,l]}+{\partial }_0\Gamma _{j[k}^pK_{l]p}+{
\Gamma }_{j[k}^p{\partial }_0K_{l]p}\right] _{(1)}  \nonumber \\
&&-4n^{[0}{m}^{j]}n^{[0}{m}^{l]}\left[ {
\partial }_0{R}_{jl}-2K_{(l}^p{\partial }
_0K_{j)p}-2N_{(0)}K_{jp}K_q^pK_l^q\right.  \nonumber \\
&&\left. +K_{jl}{\partial }_0K+K{\partial }_0K_{jl}\right] _{(1)} \nonumber \\
&& +2\{\psi_4 (l_i \Delta -m_i\bar{\delta}) N^{i~(0)}
+\psi_3 (n_i\bar{\delta}-\bar{m}_i \Delta) N^{i~(0)}\},\nonumber
\end{eqnarray}
where
\begin{eqnarray}
\psi _3 &=&-\left[ {R}_{ijkl}+2K_{i[k}K_{l]j}\right] _{(1)}
l^i{n}^j\bar{m}^k{n}^l+4\left[ K_{j[k,l]}+{
\Gamma }_{j[k}^pK_{l]p}\right] _{(1)}
(l^{[0}{n}^{j]}\bar{m}^k{n}^l-n^{[0}\bar{m}^{j]}l^k{n}^l)
\label{psi3} \\
&&-2\left[ {R}_{jl}-K_{jp}K_l^p+KK_{jl}
\right] _{(1)}
(l^{[0}{n}^{j]}\bar{m}^0{n}^{l}-l^{[0}{n}^{j]}n^0\bar{m}^{l}), 
\nonumber
\end{eqnarray}
$N_{(0)}=(g_{kerr}^{tt})^{-1/2}$ is the zeroth order lapse, 
$n^i$ in these equations should be taken to be related to that of the
original penetrating tetrad as $n^0= N_{(0)}n^0_{\rm orig}, 
n^i=n^i_{\rm orig}+N^{i~(0)}n^0$.
Latin indices run from 1 to 3, and the brackets are
computed to only first order (zeroth order excluded). The derivatives involved
in the above expressions can be computed in terms of the initial data on 
the Cauchy hypersurface as,
\begin{equation}
{\partial }_0K=N_{(0)}K_{pq}K^{pq}-{\nabla }%
^2N_{(0)},  \label{Kpunto}
\end{equation}
\begin{equation}
{\partial }_0{R}=2K^{pq}{\partial }%
_0K_{pq}+4N_{(0)}K_{pq}K_s^pK^{sq}-2K{\partial }_0K,
\label{Rpunto}
\end{equation}
\begin{eqnarray}
{\partial }_0R_{ijkl} &=&-4N_{(0)}\left\{ K_{i[k}{R}%
_{l]j}-K_{j[k}{R}_{l]i}-\frac 12{R}\left(
K_{i[k}g_{l]j}-K_{j[k}g_{l]i}\right) \right\}  \label{Rijklpunto} \\
&&\ +2g_{i[k}{\partial }_0{R}_{l]j}-2g_{j[k}{%
\partial }_0{R}_{l]i}-g_{i[k}g_{l]j}{\partial }_0%
{R}+2K_{i[k}{\partial }_0K_{l]j}-2K_{j[k}{%
\partial }_0K_{l]i},  \nonumber
\end{eqnarray}
and,
\begin{equation}
{\partial }_0K_{ij}=N_{(0)}\left[ \bar R%
_{ij}+KK_{ij}-2K_{ip}K^p{}_j-N_{(0)}^{-1}\bar \nabla _i\bar \nabla
_jN_{(0)}\right] _{(1)}.  \label{Kijpunto}
\end{equation}

We calculate the Teukolsky function $\psi$ and its time derivative $\dot \psi$, on the initial time slice, 
using the above formulae on expressions for the metric and extrinsic curvature that we 
produced before.  The algebraic expressions for the initial data of the Teukolsky function are too
large to include in this paper. Instead, we include some surface plots.

\begin{figure}
\centerline{\psfig{file=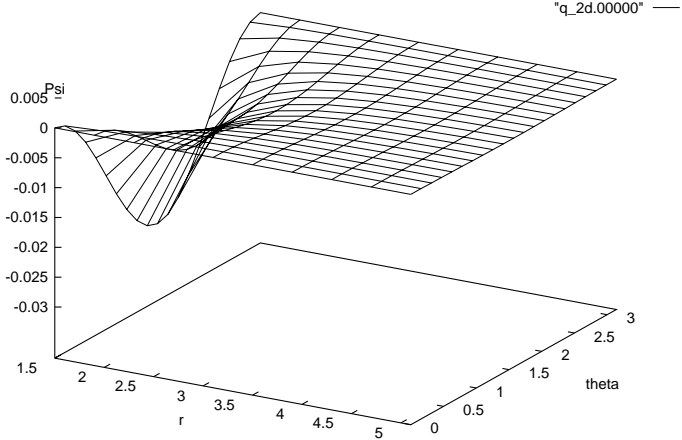,height=72mm}}
\caption{The initial Teukolsky function, as function of $r$ and $\theta$. The value
of the constant $A_2$ here is $0$.}
\end{figure}

\begin{figure}
\centerline{\psfig{file=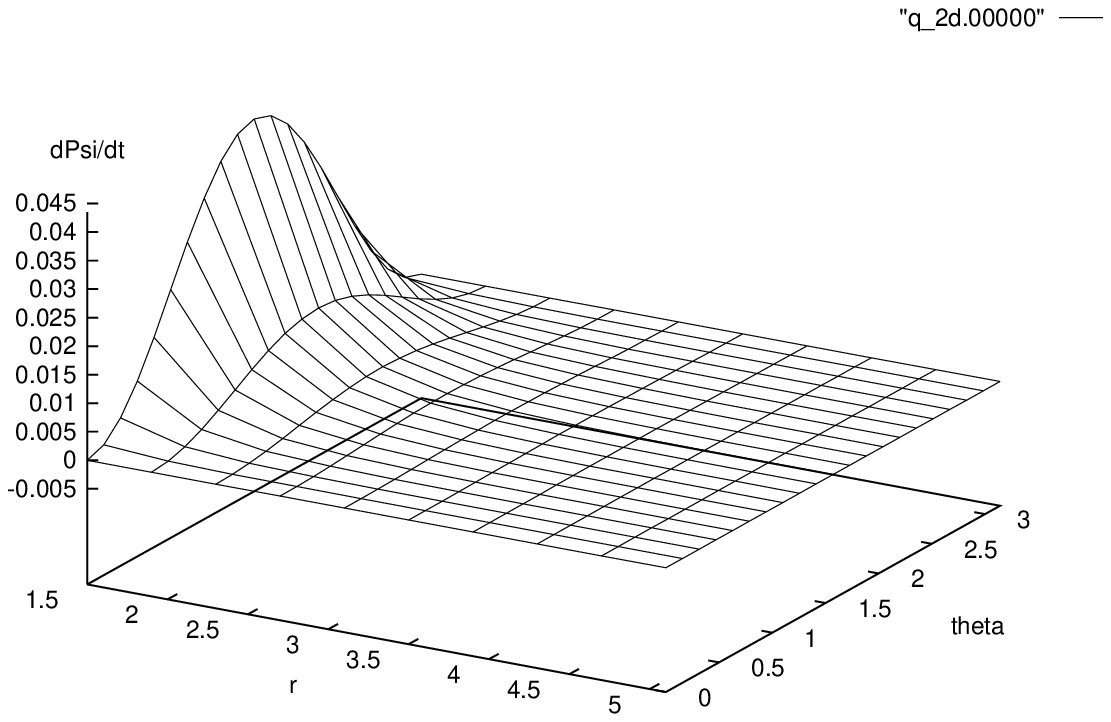,height=72mm}}
\caption{The initial time derivative of the Teukolsky function, as function of $r$ and $\theta$. 
The value of the constant $A_2$ here is $0$.}
\end{figure}

We still need to discuss the choices for the unknown constants $A_1$ and $A_2$ that appear in  
equation {\ref{Eq-vSol}} and therefore in expressions for $\psi$ and $\dot \psi$.  It turns out that in 
order for the Teukolsky function to have a proper fall-off behavior for large $r$, $A_1$ must equal $-12$.
This choice for $A_1$ also ensures that $v(r)$ vanishes at spatial infinity. 
$A_2$ on the other hand can be chosen at will. In their paper, Sarbach {\em et al.} provide a very
clear explanation regarding the meaning of $A_2$ based on the original assertion by Bishop {\em et al.} 
They show that the meaning of $A_2$ is related to the position of the apparent horizon, through a genuine 
choice (not just a gauge choice) in the initial data. 

In what follows, we shall choose ${A_1}\,=\,-12$ and use multiple values of $A_2$. In particular,
we shall use ${A_2} = 0$ and ${A_2} = 10$ for our results.

\section{Evolution}

Like we mentioned in the previous section, the initial data for the Teukolsky function obtained 
above, shall be evolved using the penetrating Teukolsky equation. This equation is best suited 
for comparisons with current KS type full numerical binary black hole codes like MAYA, because
its coefficients are regular at the horizon. This equation was derived by Campanelli {\em et al.} 
\cite{CKLPR}, and we reproduce it here (with the Kerr parameter, $a=0$),
\begin{eqnarray}
&&
(\Sigma + 2Mr){{\partial^2 \psi}\over 
{\partial t^2}} - \triangle {{\partial^2 \psi}\over 
{\partial r^2}} - 6(r - M){{\partial \psi}\over
{\partial r}}
\\
&&
-{{1}\over {\sin \vartheta}}{{\partial}\over {\partial \vartheta}} \left (
\sin \vartheta {{\partial \psi}\over {\partial \vartheta}}\right ) -{{1}\over
{\sin^2 \vartheta}}{{\partial^2 \psi}\over {\partial  \varphi^2}} 
-4Mr{{\partial^2 \psi}\over {\partial t \partial r}} 
\nonumber\\
&&
+ \left ({4i\cot\vartheta \over \sin\vartheta}
\right ) {{\partial \psi}\over {\partial \varphi}}\nonumber\\
&& 
- \left ({4r+6M}\right ) {{\partial \psi}\over {\partial
 t}} 
 + 2(2\cot^{2}\vartheta-1)\psi = 0.
\nonumber
\end{eqnarray}
To implement the penetrating Teukolsky equation numerically, we reduced it down
into a 2+1 dimensional one, using a decomposition of the Teukolsky
function into angular modes, $\psi=\Sigma \psi_{m} e^{im {\varphi}}$.
We used Lax-Wendroff technique to numerically implement these simplified
set of equations.

\begin{figure}
\centerline{\psfig{file=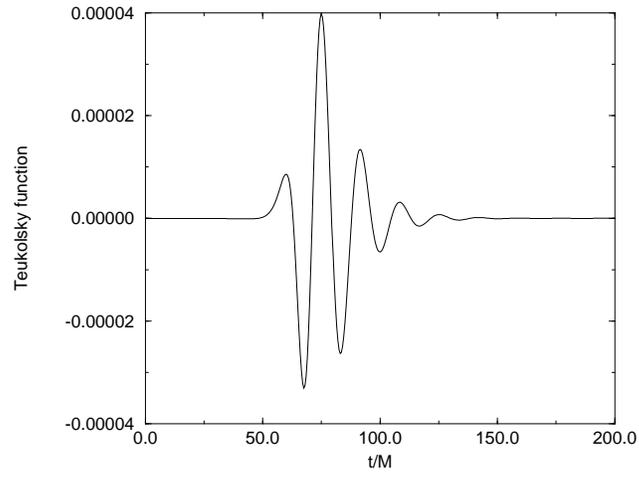,height=72mm}}
\caption{The radiated waveforms at $r=50M$ for $A_2=0$. As usual in close limit collisions, the waveform is
dominated by the fundamental quasi-normal mode.}
\label{waveform0}
\end{figure}

\begin{figure}
\centerline{\psfig{file=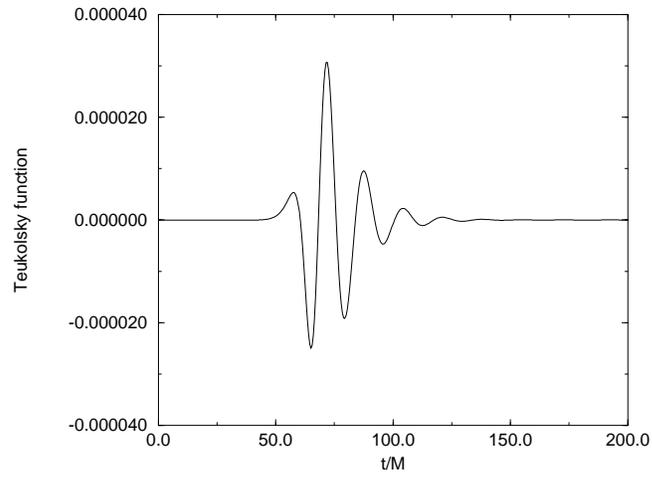,height=72mm}}
\caption{The radiated waveforms at $r=50M$ for $A_2=10$. As usual in close limit collisions, the waveform is
dominated by the fundamental quasi-normal mode.}
\label{waveform10}
\end{figure}

Results of evolutions for two equal mass KS holes are shown in the included figures. The 
inner boundary for the evolution was chosen to be inside the horizon, at $1.5M$.
 In figures \ref{waveform0} and \ref{waveform10} we show the Teukolsky function as a function 
of time, extracted at $r=50M$.  We show waveforms for two values of $A_2$. Quasi-normal ringing 
is self-evident in these figures as expected.  We also calculated the energy radiated for these evolutions, 
and we get amounts in agreement with those obtained by Sarbach {\em et al.} 
$$
E = \varepsilon ^4 M  \left( 1.31 \times 10^{-4} +
1.22\times 10^{-6}A_2 ^2 - 1.92\times 10^{-5}A_2  \right)
$$
If we pick $A_2$ such that the energy radiated 
is at a mininum (this occurs at ${A_2}= 7.87$), and we pick the conformally flat separation 
between the holes to be about $0.9M$, the value we obtain is about $10^{-5}M$, which is 
consistent with the value obtained from the evolution of Misner data (with the
same separation $L=0.9M$ or ${\mu_0}=1.5$)  using the original Teukolsky or Zerilli 
frameworks \cite{pp}.  

\section{Conclusions}

We evolved the initial data family of Bishop {\em et al.} in the
close-limit  by treating the spacetime as a single distorted Kerr--Schild 
black hole. We performed the evolution using the penetrating Teukolsky equation.

These results shall be used in the calibration of numerical codes based on the
Kerr-Schild coordinates, e.g. MAYA \cite{maya}. They are also 
in agreement with those obtained by Sarbach {\em et al.} 
from their evolution of the same initial data family using a KS type Zerilli equation. 

\section{Acknowledgements}
 
We thank Long Island University for research support and computational
facilities.  We also thank Ashish Tiwari for helping out with some of the
computational aspects of this paper, as well as proof reading it.


\end{document}